# Mechanical effects of carboxymethylcellulose binder in hard carbon electrodes

M. Anne Sawhney[a,b], Emmanuel Shittu[a,b], Ben Morgan [a], Elizabeth Sackett [a], Jenny Baker[a,b,*]

[a]  Faculty of Science and Engineering, Swansea University Bay Campus, Fabian Way, SA1 8EN, UK
[b]  SPECIFIC Innovation and Knowledge Centre, Swansea University Bay Campus, Fabian Way, SA1 8EN, UK
*   Correspondence: j.baker@swansea.ac.uk

**Abstract**

   Electrodes in sodium-ion batteries endure mechanical stress during production and application, which can damage these fragile coatings, causing performance inefficiencies and early failure. Binder material provides elasticity in electrode composites to resist fracture, but evaluating the effectiveness of binder is complicated by substrate dependency of these films, while conventional cell tests are beset by multiple electrochemical variables. This work introduces a practical low-cost indentation test to determine the elasticity of hard carbon electrodes containing standard carboxymethylcellulose binder. Using the proposed method, relative elastic moduli of hard carbon electrodes were found to be 0.079 GPa (1% binder), 0.088 GPa (2% binder), 0.105 GPa (3% binder) and 0.113 GPa (4% binder), which were validated using a computational model of film deflection to predict mechanical deformation under stress. Effects on the electrochemical performance of hard carbon anodes were also demonstrated with impedance spectroscopy and galvanostatic cycling of sodium half-cells, revealing 8-9% higher capacity retention of anodes with 4% binder compared with those containing 1% binder. These findings suggest binder content in hard carbon electrodes should be selected according to requirements for both cycle life and film flexibility during cell manufacturing.

# 1. Introduction

Demand for energy storage solutions is increasing globally, and despite rapid expansion of lithium-ion battery production, under-supply is projected to continue [1,2]. Less developed alternatives, such as sodium-ion batteries, are marketed as possible replacements for lithium-ion products in some applications [1,2]. At the time of writing, imminent large-scale commercial release of sodium-ion (Na-ion) batteries for both stationary and transport applications has been announced by multiple manufacturers [1,3]. Large scale commercial production of battery cells can encounter many problems with electrode production with manufacturing scrap rates reportedly up to 90% in the early stages of start up [4], for this reason it is important to understand the mechanical as well as electrochemical performance of these materials.

Graphite anodes used for lithium-ion cell applications are known to endure mechanical stresses related to intercalative expansion and contraction during cycling [5]–[7]. In contrast, hard carbon anodes common to Na-ion cells are reported to be more resilient to electrochemically-linked damage such as intercalation-induced exfoliation, due to the disordered nanostructure of this active material [8]. However, both graphite and hard carbon electrodes can incur fracture damage during electrode manufacturing, such as mixing [9], drying [10] and compressing/calendaring [11]. Although the commercial impact of this type of electrode damage is rarely reported, evidence from studies of lithium-ion cells confirm even minor defects caused during roll-to-roll production decreases electrochemical performance of cells [12,13]. Protection from mechanical stresses during cell cycling is provided by binder material in the electrode formula, while less evidence is available concerning the effectiveness of this component in preventing damage to dry electrode films resulting from typical manufacturing processes.

The default binder applied to graphite anodes in lithium-ion cells is sodium carboxymethylcellulose (CMC), a water-soluble long-chain carbohydrate [14]. Advantages of water-based slurry processing combined with the compatibility of CMC with hard carbon suggest this binder is also an industrial standard for Na-ion anodes [15]. The effectiveness of CMC in producing graphite anodes with microstructural integrity is well established [12], while proportionately less is reported about the mechanical integrity of hard carbon/CMC anodes in Na-ion cells [16].

A 2021 review [17] of electrode binders in Na-ion research literature listed studies comparing CMC to alternatives, though the diverse range of different active materials and characterisation methods employed by each laboratory obstructs comparisons between studies of binder effectiveness. The same year, Gond et al. [18] proposed sodium lignosulphate as an alternative binder to CMC for hard carbon anodes, simultaneously revealing effects of binder selection on the composition and thickness of the solid-electrolyte interphase (SEI). More recently, Cao et al. [19] demonstrated MXene as an electronically conductive binder for hard carbon anodes, mitigating an inherent disadvantage of standard binders such as CMC. A trend toward "binder-free" anodes for Na-ion is also apparent in literature [20]–[22], seemingly rendering this component of slurries obsolete. However, the absence of quantitative mechanical testing of Na-ion electrodes proposed in these publications prevents comparison with current standards or evaluation of reproducibility, which are prerequisites of scalability to commercial production.

Several techniques to test mechanical properties of electrode films have already been applied in lithium-ion cell research. Dry methods, such as bend tests [23] and peel-off tests [24], generally assess adhesion of the electrode films to current collector substrate rather than bulk electrode properties. Novel experimental approaches, such as the visible bending of a cell suspended in electrolyte, allow direct correlation between mechanical deformation and the electrochemical state of the anode [25]. Due to the fragility of electrode films and proportionally higher elasticity of underlying foil substrates, quantifying flexibility in dry carbonaceous anodes is impractical using traditional tensile testing [26]. Tensile testing of these electrodes relies on a rule of mixtures theory [27], which assumes a portion of the measured force response can be attributed to the coating, despite evidence showing this practice overestimates stress in the film [28].

In contrast, nanoscale mechanical characterisation equipment is better suited to assessing elastic properties at the surface of substrate-dependent films. Nanoindentation [29,30] and nano-cantilever methods [31]–[33] have been employed in analysis of electrode elasticity, with increasing application in the study of electrodes for lithium-ion batteries [34,35]. However, the high costs and specialist training associated with this method is a barrier to more widespread application, particularly outside the research environment.

Before the development of nanoindentation techniques, larger-scale indentation was industrially applied for the quantification of hardness in metals [36]. Studies by Stilwell and Tabor on the indentation of metals revealed a relation between unloading-displacement curves, Poisson's ratios and elastic moduli [37]. The following work by Bulychev, Alekhin, Shorshorov et al. established a method for calculating elastic modulus of an indented sample based on the initial slope of the unloading curve, representing stiffness under elastic deformation [37].

In this work, indentation theory is used to evaluate CMC binder effectiveness in hard carbon composite films typical of Na-ion anodes by quantifying bulk elastic moduli of thick solid samples. A proposed millimetre-scale indentation method resolves the difficulty of conventional tensile testing techniques for fragile coated electrode films, while employing low-resource-intensive procedures relative to nanoindentation. The slope of unload-displacement curves was used to quantify the relative elastic moduli of hard carbon electrodes, which were validated as a measure of binder effectiveness by analysing the bending deformation of films loaded experimentally and computationally using Finite Element Analysis (FEA). Finally, the electrochemical performance of hard carbon anodes containing 1 to 4 % CMC was determined using galvanostatic cycling of half-cells and impedance spectroscopy.

## 2. Methods

### 2.1 Experimental investigation of hard carbon electrodes

#### 2.1.1 Preparation of hard carbon electrode test samples and films

Hard carbon electrode films and thick samples were made (process details available by video at [38]) from 9 µm Kuranode type 2 hard carbon (Kuraray), C65 carbon black (Imerys) and CMC (Merck, 0.9 degree of substitution, 250 kDa) in solid proportions of 97:2:1, 96:2:2, 95:2:3 and 94:2:4. Slurries were mixed into deionized water (Millipore) in the order of CMC, C65 then hard carbon. Suspensions were agitated at 3500 revolutions per minute with a Speedmixer (DAC 150.1 FVK-K, Synergy Devices Ltd), while initial (CMC solution) and final steps (pre-spreading) consisted of magnetic stirring between 200 and 600 revolutions per minute, before applying onto 15 µm thick aluminium current collector (H18 alloy, MTI Corp).

Electrode thicknesses were tested by micrometre at three points on each film (Mitutoyo, 0 – 25 mm ± 0.001 mm). Thicknesses of current collector foil were verified by the same method, then subtracted from average electrode thickness to determine the thickness of each hard carbon coating. Thicker hard carbon electrode film samples for proposed indentation tests (described in section 2.1.2) were made by pouring slurry into slabs > 5 cm diameter with thickness over 2 mm. Films for deflection and electrochemical tests (described in sections 2.1.3 and 2.1.4) were tape-cast with a glass bar supported across two layers of translucent tape (Scotch® Magic™ Greener Choice) and dried in a humidity-controlled cleanroom for a minimum of 24 hours.

*2.1.2   Indentation tests*

The proposed millimetre-scale indentation test method was performed using a digital force gauge (Mark-10 Series 7, 10 N maximum). A conical indenter (Mark-10, tip diameter ≈ 100 µm) attached to the force gauge was lowered into hard carbon sample slabs (loading) then raised (unloading) at incremental depths (Fig. 1a). At each depth quantified by a generic digital micrometre, the corresponding displacement and force values in Newtons after 1 minute of relaxation were recorded manually. A minimum of two indents were performed on different days for each sample type, with maximum depth of all indents between 0.17 mm and 0.20 mm. To verify the portion of the unloading curve used for analysis represented an elastic region of deformation, for several indentations the same point was re-loaded three times in succession, confirming load vs displacement curves overlapped in this region.

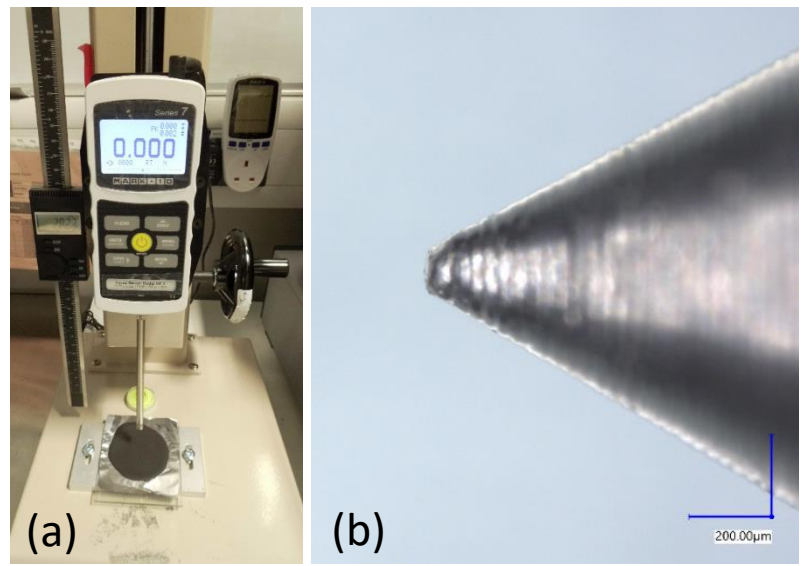

**Fig. 1.** Indentation test setup and magnified conical tip used for indentation tests.

The dimensions of the indenter cone were determined using a Keyence VHX-7000 optical microscope. The projected area, *A*, of contact between the blunt conical indenter and samples at maximum loading was estimated by calculating the cone base area of an equivalent truncated cone with tip radius 50 µm and length equal to maximum indentation depth (Fig. 1b).

*2.1.3   Film Deflection tests*

To demonstrate the predictive capability of the proposed model and confirm the relevance of relative moduli obtained by indentation, experimental deformation tests were performed on films matching those simulated computationally. Hard carbon films spread onto aluminium foil with dry thickness of 0.06 to 0.071 mm were cut and mounted to a custom sample holder matching the dimensions of the model (65 mm x 90 mm) and subjected to mid-span deflection (10 mm wide) of 2 mm depth using a Hounsfield universal mechanical tester (Fig. 2). Three tests were performed on films of each binder loading, with force measurements recorded for comparison with simulation results.

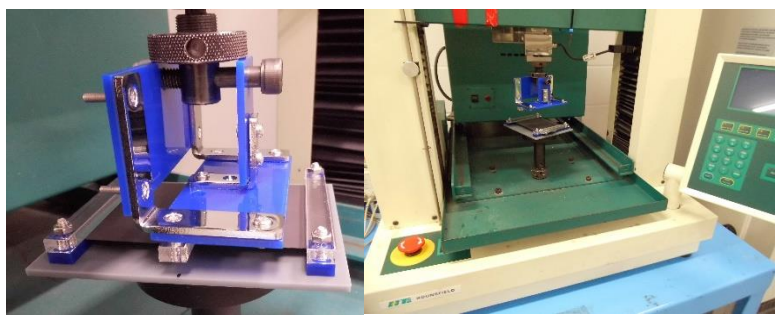

**Fig. 2**. Custom setup used for film deflection tests.

*2.1.4  Electrochemical tests*

Electrodes used for electrochemical tests were baked at 120°C in a vacuum antechamber (MBRAUN) for a minimum of 8 hours before transferring into an argon glovebox. Hard carbon electrodes, polypropylene separator (Celguard 2500) and glass fibre (Ohaus) were cut to 16 mm diameter discs and pre-soaked with electrolyte composed of 1 M $NaClO_4$ (Merck) in ethylene carbonate (99%+, Thermo Fisher Scientific) and dimethyl carbonate (99%+, Thermo Fisher Scientific) at equal proportions by weight. CR2032 coin cells (MTI Corp) applying one aluminium spacer (MTI Corp) and one wave spring (MTI Corp) were constructed with a sodium metal counter electrode (Merck) rolled out with a pasta maker and cut to 15 mm diameter. A minimum of three Na-ion half cells of each type were made and tested on two different days to account for the effects of inter-day differences on repeatability.

Coin cells were placed into generic coin cell holders for galvanostatic cycling with a Maccor 4300 battery tester and electrochemical impedance spectroscopy (EIS) with an Ivium CompactStat. Before and after cell cycling, two-electrode EIS was performed between 100 kHz and 0.1 Hz at 0.01 V amplitude centred at open circuit potential determined with a 15 s measurement. Three cells of each type (1%, 2%, 3% and 4% CMC) were cycled 5.5 times between 2.0 V and 0.005 V at 300 µA, corresponding to 30 mA $g^{-1}$ for a typical anode active material loading of 5 mg $cm^{-2}$. To intensively stress electrodes, further cycling was performed on one cell of each type applying 30 mA (15 mA $cm^{-2}$) for five cycles. To test electrode resilience to multiple sources of stress, cycling was performed on remaining cells by applying 3 mA (1.5 mA cm$^{-2}$) for 500 cycles, after storing at sodiated state for 4 weeks either before or after the first 5.5 cycles.

## 2.2 Calculating relative elastic modulus

The slope obtained at the start of the unloading curve was equated to stiffness for calculating the relative elastic modulus of the hard carbon film ($E_r$) using equations (1) to (3).

$$\text{measured stiffness,} \quad S = (2 \cdot E_r \cdot \sqrt{A}] / \sqrt{\pi} \quad (1)$$

$$\text{displacement at the indented surface,} \quad h_s = (0.72 \cdot P_{max}) / S \quad (2)$$

$$\text{loaded contact depth,} \quad h_c = h_{max} - h_s \quad (3)$$

where $E_r$ is relative modulus combining both sample and indenter, $A$ is the projected area of the indentation, $P_{max}$ is peak indentation load, and $h_{max}$ is indenter displacement at peak indentation load. Although Poisson's ratio values of hard carbon films have not been established in literature, the elastic modulus of the steel indenter is much greater than the expected values of the electrode film, therefore the relative modulus was anticipated to be a close approximation of the true elastic modulus of the bulk composite material.

## 2.3 Deformation model

The bending deformation analysis was completed on a research license of Ansys 2023R1 using a Static Structural analysis system. The geometry was defined with a length and width of 90 mm and 65 mm. The thickness of the aluminium current collector was measured as 0.015 mm, and from a series of measurements, the hard carbon film average thickness was taken as 0.07 mm. With an effective total thickness of 0.085 mm, the ratio of thickness to length is <1/1000 so plate theory can be employed to simplify the model. Two surfaces were created with the area dimensions in the XY-Plane, and the thicknesses defined as a property of each surface in the Z-Plane. The upper hard carbon film surface had its top face split with a 10 mm wide strip for the application of the displacement boundary condition, applying 2 mm in the negative Z-direction (Fig. 3). Fixed support constraints were added to the short edges of the sheets. Two variations of this constraint were tested: just the two hard carbon film edges, and then all four edges of both sheets. Prior to running the models, it was undetermined as to which constraint would be most suitable. Both variations produced valid sets of data and will be presented in the results.

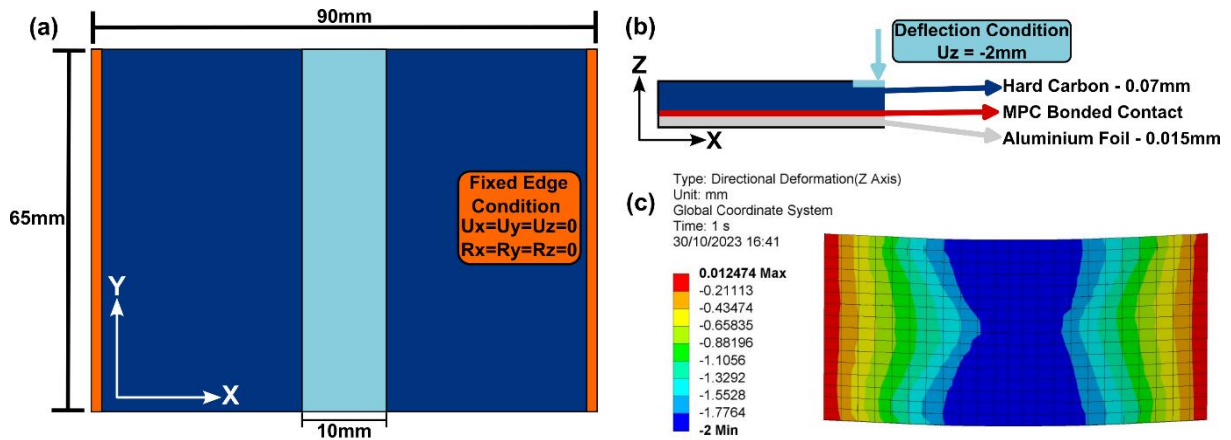

**Fig. 3.** Hard carbon film modelling, (a) diagram of model showing boundary condition locations, (b) arrangement of surfaces with contact and displacement condition, (c) a displacement plot from ANSYS.

Isotropic elastic material models were used, with relative moduli values derived from the presented experimental work. The value for Poisson's ratio of hard carbon films was taken from literature (0.3; [6,39]). A sensitivity study was completed to determine the suitability of this value, testing values between 0.25-0.35 on the force reaction result. The results of this study established the Poisson's ratio to be insignificant with a correlation of 0.103 and R2 contribution of 0.01. To capture the behaviour of the hard carbon film adhering to the current collector, a bonded contact region was defined between the two surfaces and the contact formulation set to a Multi-Point Constraint (MPC). This prevents any separation and stiffens the surface contacts by replacing the springs between contact nodes with rigid beams. A Sparse Matrix Solver was used with settings adjusted to turn off weak spring formulation and turn on large deflection due to the displacement boundary condition. Both sheets were meshed with the 4-node SHELL181 element that has 6 degrees of freedom at each node. This element is suitable to modelling sandwich construction and utilises first-order shear deformation theory. A mesh study was completed, and the model determined to be mesh dependent. From benchmarking against experimental data, an element size of 4 mm was determined to be suitable to provide accurate results. The total number of elements and nodes came to 736 and 816.

## 3. Results and discussion

*3.1 Electrode analysis by indentation*

The aim is to quantify the binder's effect on electrode elasticity by calculating relative elastic modulus for a range of binder content. Increasing the proportion of binder in an electrode formula is known to alter microstructure through additional flexible inter-particle connections [40], while consequent macrostructural effects such as film elasticity have not been established for hard carbon films. In this work, hard carbon electrodes were tested with a multi-purpose force gauge and a millimetre-scale indenter as a simple low-cost alternative to nanoindentation.

Using results from the proposed indentation tests, unload-displacement curves (Fig. 4a) were plotted for each hard carbon film. During unloading, high deviation of force and displacement values within sample types reflects changing conditions, such as temperature, between repeats deliberately performed on different days. However, it is the slope at the start of the unloading curve that represents material stiffness in the elastic region, and despite the observed inter-test variations in measured values, a trend of increasing slope with increasing CMC was observed (inset to Fig. 4a).

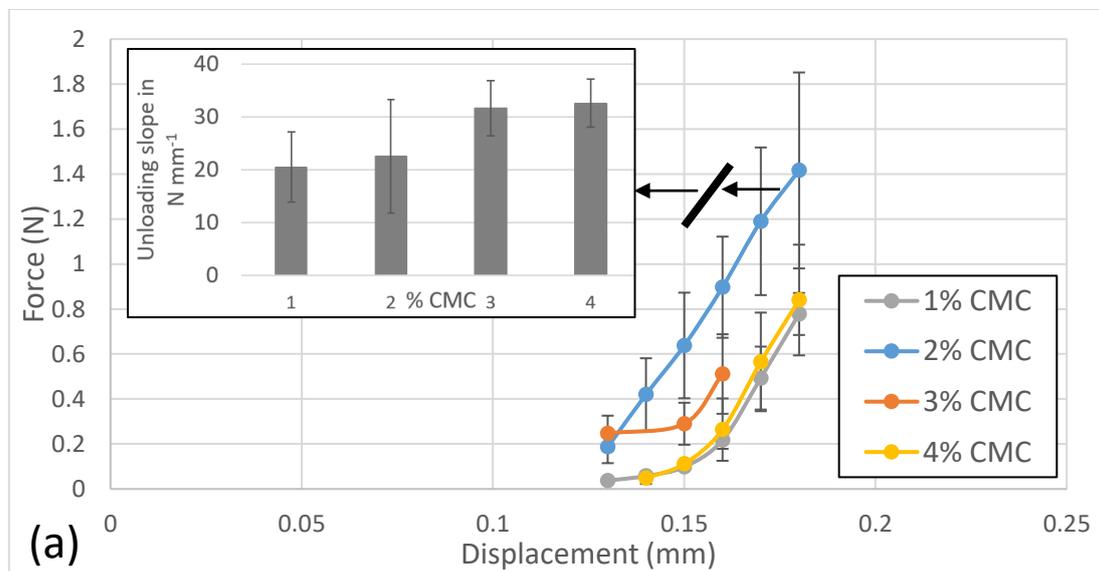

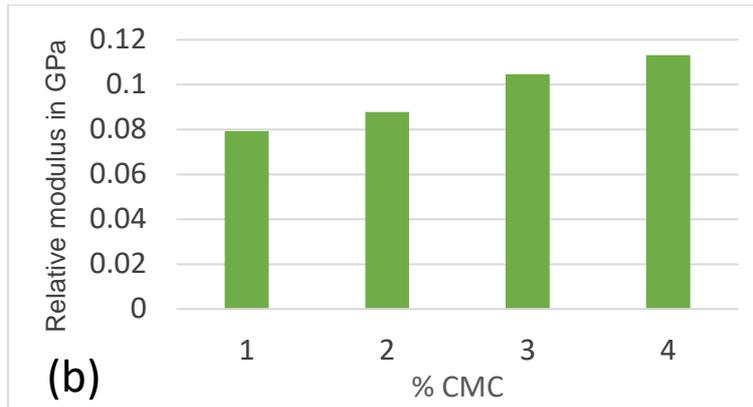

**Fig. 2.** Indentation unloading (a) curves showing mean values and initial slope on inset, and (b) corresponding relative elastic moduli; bars represent standard deviation.

The slopes of the initial data points of the unloading curves were applied as the stiffness of each hard carbon film in calculating relative elastic moduli (using equation 1) according to binder content (Fig. 4c). Across the range of binder loading tested, the stiffness and relative modulus of hard carbon electrode samples increased proportionally with CMC binder content. To validate the obtained relative moduli as values representative of bulk dry film flexibility, these values were applied experimentally and computationally to 65 mm wide electrode sheets, which were subjected to bending within the elastic region to simulate forces on a dry electrode during cell manufacturing.

*3.2 Bending deformation analysis*

To assess the relevance of relative elastic moduli values obtained by indentation, a deflection test was performed on models of hard carbon films on current collector foil (described in Section 2.3) with results compared to data from a matching experiment (described in Section 2.1.3).

Simulations were completed for each binder percentage, and force reaction values collected from the displacement boundary condition (Fig. 5). The transparent lines are experimental data from which the model material properties were derived. The grouping of these experimental sets is reasonably tight and provides a suitable benchmark to determine the efficacy of the model. As mentioned in section 2.3, two variations of the fixed boundary condition were assessed. On the graph, the solid line represents the setup where only the two hard carbon film edges were fixed (denoted as 2E), and the dotted line represents where both surfaces had the edges fixed (denoted as 4E). The final force reaction value for the 4E condition aligns with the upper bounds of the experimental data of all binder percentages, except for 1%. This condition also displays noticeably different behaviour in the shape of the data. When compared to the 2E condition, the 4E data demonstrates more compliancy in early stages of loading, before becoming stiffer and producing a larger final reaction force. The force convergence data indicates that compliant behaviour at lower displacement is a result of the solver attempting to correct for an over-constrained model.

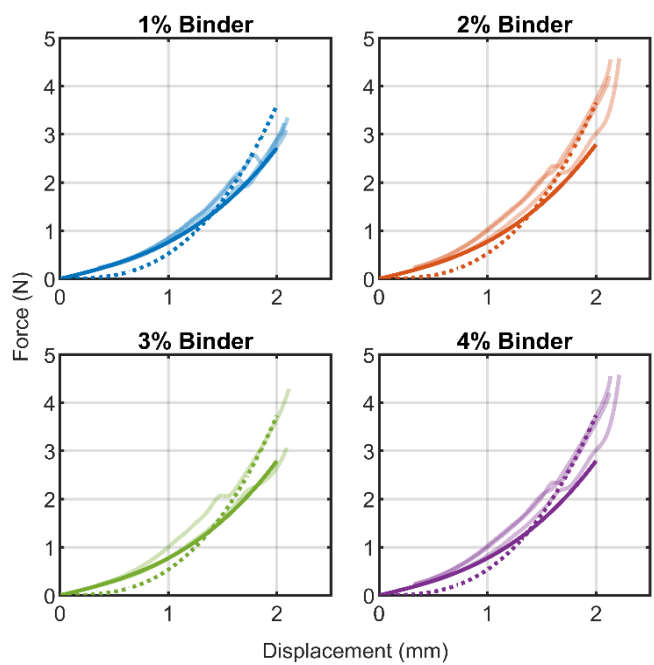

**Fig. 5.** Graphs of force against displacement for all four binder percentages. Transparent lines: experimental data, dotted line: 4-edge fixed condition (4E), solid line: modelled film edge fixed condition (2E).

The 2E conditions demonstrates a behaviour that is more consistent with that of the experimental data. This is most notable with the 3% binder, where the simulation data is almost an exact match for the lower experimental data set. A high similarity can also be seen with the 1% binder model, where it should also be noted that the experimental data has the tightest repeatability. The 2E condition model slightly underpredicts the maximums for the 2% and 4%. When comparing the experiment average force values at 2 mm displacement to the simulation results, the 1% and 3% model see a difference of 6.4-9.5%, whilst the 2% and 4% models see a difference of 19.5-20%. It is possible that the differences in values for these specific binder contents are due to some inconsistencies in the sample preparation for the experimental data. The closeness in data for the 1% and 3% binder models indicate that the model can analyse the bending deformation of the electrode to a sufficient degree. Finally, model demonstrates that increasing the binder content also increases the bending stiffness of the electrode from 1.33 N mm$^{-1}$ at 1%, to 1.45 N mm$^{-1}$ at 4%.

*3.3 Electrochemical performance*

To determine whether differences in CMC binder content affected measurable performance characteristics of hard carbon anodes in a typical cell, sodium metal half-cells were constructed and galvanostatically cycled, with impedance spectroscopy performed at periodic intervals.

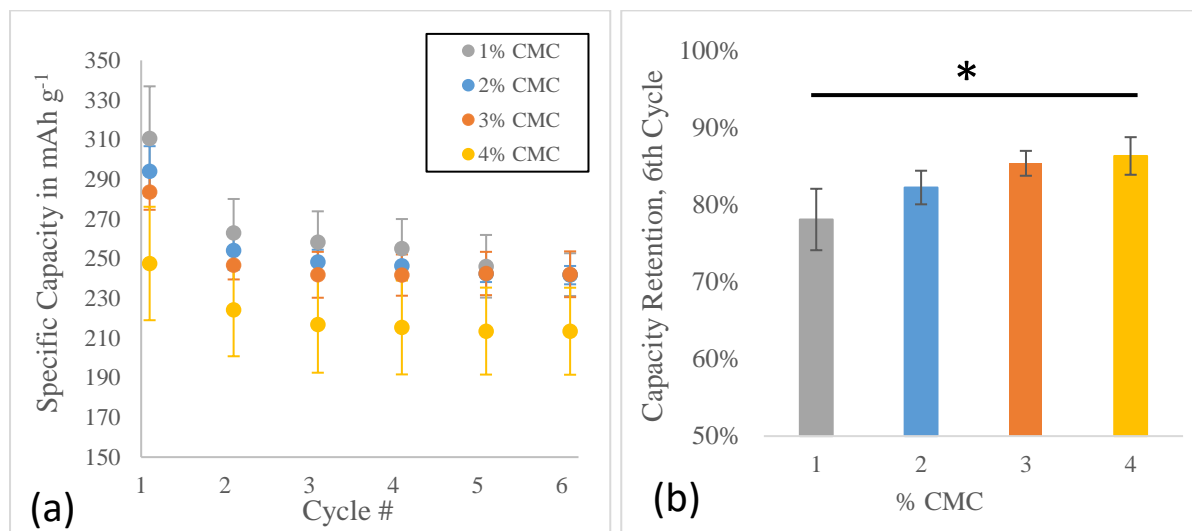

**Fig. 6.** Specific capacity (a) and capacity retention (b) of half-cells containing anodes with different binder loadings, after first cycles at 0.15 mA cm$^{-2}$; bars represent standard deviation, *p=0.048.

Specific capacity in the first cycle after formation was inversely proportional to binder content, which was expected due to additional binder blocking some of the reactive surface on hard carbon particles (Fig. 6a). This advantage of lower binder content dwindles after the first cycle, when the formation of an SEI layer covers active material surfaces at the reactive interface. After the fifth sodiation, specific capacity is decreased by over 20% in anodes with 1% CMC, while decreases are proportionally lower with each additional percent of binder. This loss in capacity is significantly different to anodes containing 4% CMC after just 6 cycles (Fig. 6b).

When the half-cells were subsequently cycled at a tenfold higher current rate (1.5 mA cm$^{-2}$) and hundred-fold current rate (15 mA cm$^{-2}$), the amount of CMC did not affect specific capacity of cycled half-cells (Fig. 7). Although increasing current was hypothesised to increase mechanical stresses in the electrode, faster sodiation and desodiation might instead decrease the total volumetric changes related to intercalation; this would match evidence from cycling graphite electrodes, in which lithiation efficiency was found to be inversely proportional to charge and discharge rates [41].

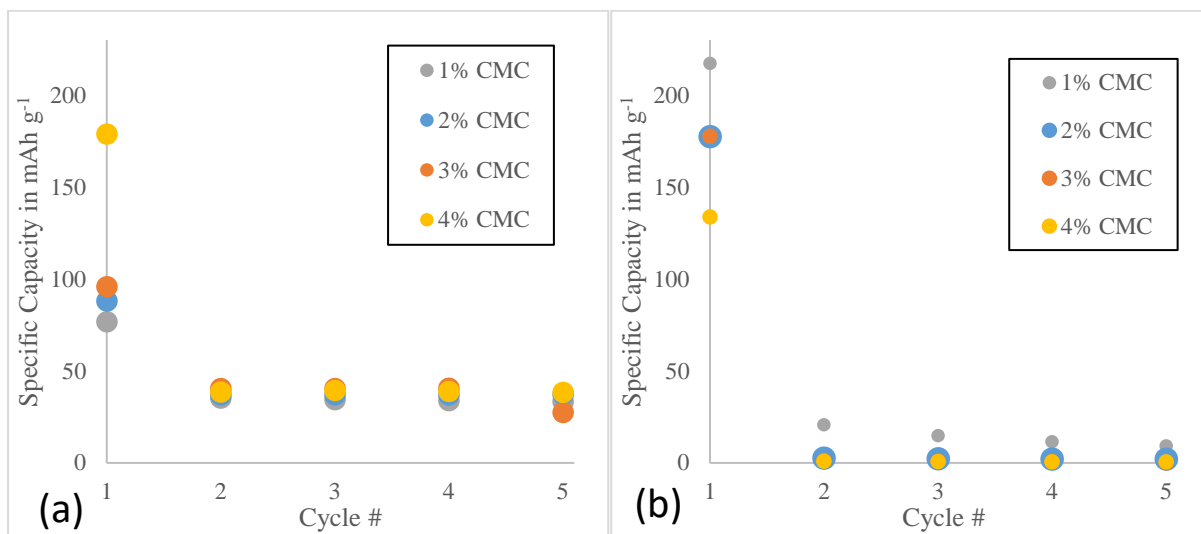

**Fig. 3.** Specific capacity of half-cells containing anodes with different binder loadings, further cycled at (a) 1.5 mA cm$^{-2}$ and (b) 15 mA cm$^{-2}$.

In contrast, crack propagation and increases in impedance between 300 and 500 cycles has been directly related to binder effectiveness in graphite anodes for Li-ion cells [42]. To observe whether this also occurs for hard carbon electrodes, two cells of each binder loading were further cycled at 1.5 mA cm$^{-2}$ for 500 cycles.

After the first five cycles at moderate current ending in anode sodiation, real and imaginary impedances of cells with more binder were slightly higher than those with less CMC (Fig. 8a). This effect of binder on impedance could be anticipated, since CMC is not electronically conductive, therefore higher quantities of CMC would be expected to decrease current carrying efficiency of an electrode. After 100 cycles at a higher current, impedances did not substantially increase, while subtle changes in spectra could be attributed to several electrochemical variables at either electrode (Fig. 8b). As cells were cycled further, impedance gradually increased, but no trend was observed according to binder proportion (Fig. 8c). While longer cycling might be required to reveal the threshold of electro-mechanical breakdown, for the procedures selected in this work, hard carbon anodes containing only 1% CMC did not increase in impedance more than those containing 4% CMC.

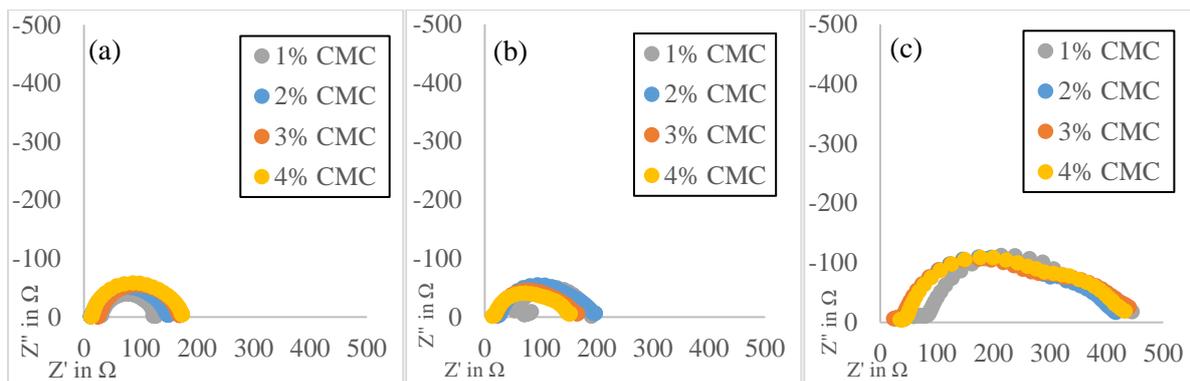

**Fig. 8.** Impedance spectra of half-cells containing anodes with different binder loadings, after (a) initial 5 cycles, (b) 100 cycles and (c) 500 cycles; averaged values for each loading.

Since impedance values across a broad frequency range did not vary with CMC content, increasing binder was not found to decrease electrical inefficiencies, which would be expected consequences of either crack propagation or delamination. Although comparing results between studies is complicated by differences in cycling parameters, the observed contrast with reported results for graphite electrodes might originate from the unique properties of hard carbon. Evidence from analysis of cylindrical Na-ion cells containing hard carbon anodes with PVDF as binder [43] suggested mechanical degradation had less impact on capacity retention than electrochemical factors, with the latter causing higher impact when ageing cells at high state of charge. Impedance was not found to be irreversibly affected after storing cells described herein while at sodiated state, therefore performance degradation caused by this phenomenon could vary with different test parameters, or it might be specific to the binder material. PVDF binder was shown to expand and to experience decreased elastic modulus when wetted with lithium-ion electrolyte [44], while this phenomenon has not been reported with CMC binder, which can be mechanically tuned through association with conductive additive particles [45].

As anticipated, stressing cells after storage while sodiated and 500 cycles at high relative current caused decreases in capacity at every binder loading (Fig. 9). While increasing current by a factor of 10 approximately halved capacity retention (shown in Figs. 6a and 7a), further losses of capacity from cycle 2 to cycle 500 were greatest with 1% CMC (62.4%) and lowest with 4% CMC (34.2%). Although cycling performance is not directly proportional to the relative elastic moduli determined for the hard carbon electrodes tested, higher binder content did increase relative elastic moduli in electrode films, which could be indirectly linked to retention of specific capacity after cycling. Since capacity retention is dependent on multiple electrochemical and mechanical variables, any differences observed between samples should be interpreted as the sum of more than one phenomenon.

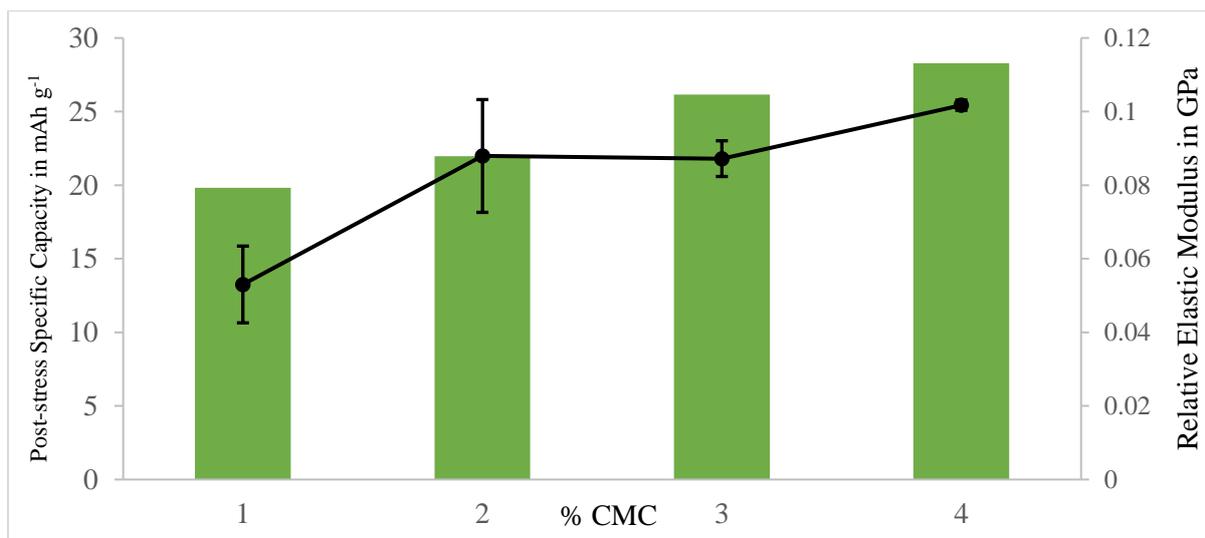

**Fig. 9.** Specific capacity after 500 cycles at 1.5 mA cm$^{-2}$ (black line) and elastic moduli as determined in this work (green columns) relative to content of CMC in hard carbon electrodes. Bars represent minimum and maximum values of plotted mean from two cells of each type.

## 4. Conclusions

In this paper, hard carbon anodes with varying proportions of CMC binder were tested to assess the effect of this common material component on electrode mechanical and electrochemical performance. Indentation was used on dry electrode samples to overcome obstacles to tensile testing techniques with films coated onto current collector foil, followed by employing conventional indentation theory and the fundamental laws of mechanics to calculate the relative elastic modulus of hard carbon electrode samples. Values of relative moduli obtained using the proposed method on standard mechanical test equipment (0.08 GPa to 0.11 GPa) demonstrated hard carbon film elastic modulus increases when CMC binder content is increased from 1% to 4% of dry weight.

The utility of relative moduli values obtained with millimetre-scale indentation were subsequently validated with an FEA-based analytical approach modelling bending deformation of electrodes on an aluminium foil substrate. The response of modelled films to bending forces matched well with results of experimental deflection of hard carbon films at all binder loadings, demonstrating the relevance of relative moduli to predicting elastic behaviour of electrode sheets. When combined with force-displacement measurements larger than particle scale, the model proposed offers a practical low-resource tool for design and optimisation of binder loading for target dry electrode elasticity.

Electrochemical results from half-cells were not directly dependent on elastic properties of hard carbon anodes, but some effects could be repeatably linked to binder proportion. Increasing CMC in the electrode formula between 1% and 4% of dry weight did not improve resilience of hard carbon electrodes to high charge/discharge rates, and after initial cycling, binder content in this range did not relate to impedance. Although initial capacity was highest in anodes with the least binder, capacity retention after SEI formation was significantly improved in anodes with more binder; after 500 cycles, specific capacity of electrodes with 4% binder was nearly double that of electrodes with 1% CMC, though this might have been an indirect result of electrochemical factors, such as electrochemically-inactive binder displacing an unstable SEI. These findings suggest carefully tuning binder content is critical to maximising hard carbon anode performance, particularly for applications requiring long cycling life.

Mechanical testing of dry electrode coatings could therefore be advantageous in optimising hard carbon electrode formulas for different cell packages and diverse user applications. Techniques such as the indentation proposed herein can be used to quantify bulk elasticity of coatings, distinct from adhesion of the film to substrate, which are both important properties to control in manufacturing hard carbon electrodes while minimising defects and scrap waste. As a complement to performance testing through electrochemical procedures, routine dry mechanical testing and computational modelling would inexpensively provide valuable and predictive information towards efficient electrode formulation and design.

## CRediT authorship contribution statement

**M. Anne Sawhney**: Conceptualization; Data curation; Formal analysis; Investigation; Methodology; Resources; Validation; Visualization; Roles/Writing - original draft; Writing - review & editing; **Emmanuel Shittu**: Formal analysis; Investigation; Methodology; Resources; Software; Visualization; Roles/Writing - original draft; **Ben Morgan**: Formal analysis; Investigation; Methodology; Software; Validation; Visualization; Roles/Writing - original draft; **Elisabeth Sackett**: Formal analysis; Investigation; Methodology; Resources; Validation; Roles/Writing - original draft; **Jenny Baker**: Funding acquisition; Project administration; Resources; Supervision; Writing - review & editing.

## Funding

This work was supported by the Engineering and Physical Sciences Research Council (EPSRC) through ECR Fellowship NoRESt (EP/ S03711X/1) and SPECIFIC Innovation and Knowledge Centre (EP/ N020863/1 and EP/P030831/1). For the purpose of open access, the authors have applied a Creative Commons Attribution (CC BY) licence to any Author Accepted Manuscript version arising.

## Acknowledgements

The authors would like to acknowledge Ben Clifford and the Welsh Centre for Printing and Coating for access to their facilities and equipment to perform this work. The authors would also like to thank Chris Batchelor at the SPECIFIC Pilot Manufacturing Research Centre for his assistance obtaining microscopy images.